\documentclass[aps,twocolumn,showpacs,amsmath,amssymb,floatfix,superscriptaddress]{revtex4}

\usepackage{graphicx}
\usepackage{dcolumn}
\usepackage{bm}
\usepackage{latexsym}
\usepackage{latexsym,graphics}

\newcommand{\be}{\begin{equation}}
\newcommand{\ee}{\end{equation}}

\def\Google{{\em Google\/}}

\begin{document}

\title{How Famous is a Scientist?  --- Famous to Those Who Know Us.}
\author{James P.~Bagrow}
\affiliation{Department of Physics, Clarkson University,
Potsdam NY 13699-5820}
\author{Hern\'an D.~Rozenfeld} 
\affiliation{Department of Physics, Clarkson University,
Potsdam NY 13699-5820}
\author{Erik M.~Bollt}
\affiliation{Department of Math and Computer Science, Clarkson University, Potsdam, NY
13699-5805}
\affiliation{Department of Physics, Clarkson University,
Potsdam NY 13699-5820}
\author{Daniel ben-Avraham}
\affiliation{Department of Physics, Clarkson University,
Potsdam NY 13699-5820}

\date{\today}

\begin{abstract}  
Following a recent idea, to measure fame by the number of \Google\ hits found in a search on the WWW, we study the relation between fame (\Google\ hits) and merit (number of papers posted on an electronic archive) for a random group of scientists in condensed matter and statistical physics.  Our findings show  that fame and merit in science are linearly related, and that the probability distribution for a certain level of fame falls off exponentially.  This is in sharp contrast with the original findings about WW II ace pilots, for which fame is exponentially related to merit (number of downed planes), and the probability of fame decays in power-law fashion.  Other groups in our study show similar patterns of fame as for ace pilots.
\end{abstract} 

\maketitle


The concept of fame is socially and economically important to many people, and the organizations to which they belong.  However, it is not a well defined concept, since each person has their own idea of what it means to be famous, including perhaps: recognizable to the ``common" person on the street (but how do we define the common person?), being on television, appearing frequently in newspapers and in other media.
Recently, researchers at UCLA, Simkin and Roychowdhury \cite{Vwani} performed an empirical study in which they catalogued the fame of World War I pilot  ``Aces."  For reasons of specificity in measurement, they chose an interesting definition of fame:  {\em the number of hits a search for a person's name garners in the \Google\ search engine\/}.  In this view, our fame is taken to be how well linked we are in what has quickly become a most popular medium --- the World Wide Web --- and is related to the number of webpages that mention us, as measured by the PageRank system that is behind  \Google's popular search engine.  This is an ingenious idea, in that it provides an inexpensive measurement of social impact, by enlisting powerful computer resources, freely available to all, to perform what would otherwise be an expensive social study.  However, we mention that in practice even this measurement is hard to make completely precise without a great deal of effort, due to difficulty in separating coincidences in popular names (doubling and tripling a person's fame), and also possibly missing a person's full fame due to too restrictive a search.  Nonetheless, in this study, it is precisely this \Google-hits measure that we adapt, as carefully specified below.

The purpose of this communication is to explore whether there is a difference between relative {\em fame\/} and {\em achievement\/} (merit) in science, as compared to the findings for ace pilots.  In the UCLA study of fighter pilots~\cite{Vwani}, a pilot's achievement was measured by how many enemy planes the pilot had downed; was he an ace?  It was found that fame  increases exponentially with achievement, while the distribution of fame falls off algebraically, nearly as $({\rm fame})^{-2}$.  A model mechanism behind these findings was presented, describing the social context of fame within a random graph.  We have catalogued similar measurements of fame and achievement for scientists working in the area of condensed matter or statistical physics.  We find dramatically different behavior, 
with fame increasing {\em linearly\/} with achievement, and its probability falling off {\em exponentially\/}.
A simple argument to account for these facts suggests that in difference to ace pilots, that enjoy public renown, scientist are well known mostly within their own community and do not truly reach real fame.
Preliminary studies of other groups of people reveal similar fame patterns as for ace pilots.

We begin with a brief summary of the findings in~\cite{Vwani}.  The distribution of fame $F$ (as measured by \Google\ hits) falls off roughly as an inverse square, 
\begin{equation}
\label{fame_aces}
P(F)\sim F^{-\gamma};\qquad \gamma\approx 2,
\end{equation}
while it rises exponentially with achievement $A$ (number of downed planes),
\begin{equation}\label{FAV}
 F(A)\sim e^{\beta A}.
 \end{equation}
The two relations imply that achievement ought to be exponentially distributed:
\begin{equation}
P(A)\sim e^{-\alpha A};\qquad \alpha=\beta(\gamma-1)\approx\beta.
\end{equation}
This is indeed confirmed from independent measurements.
 
\begin{figure}
\includegraphics[width=0.4\textwidth]{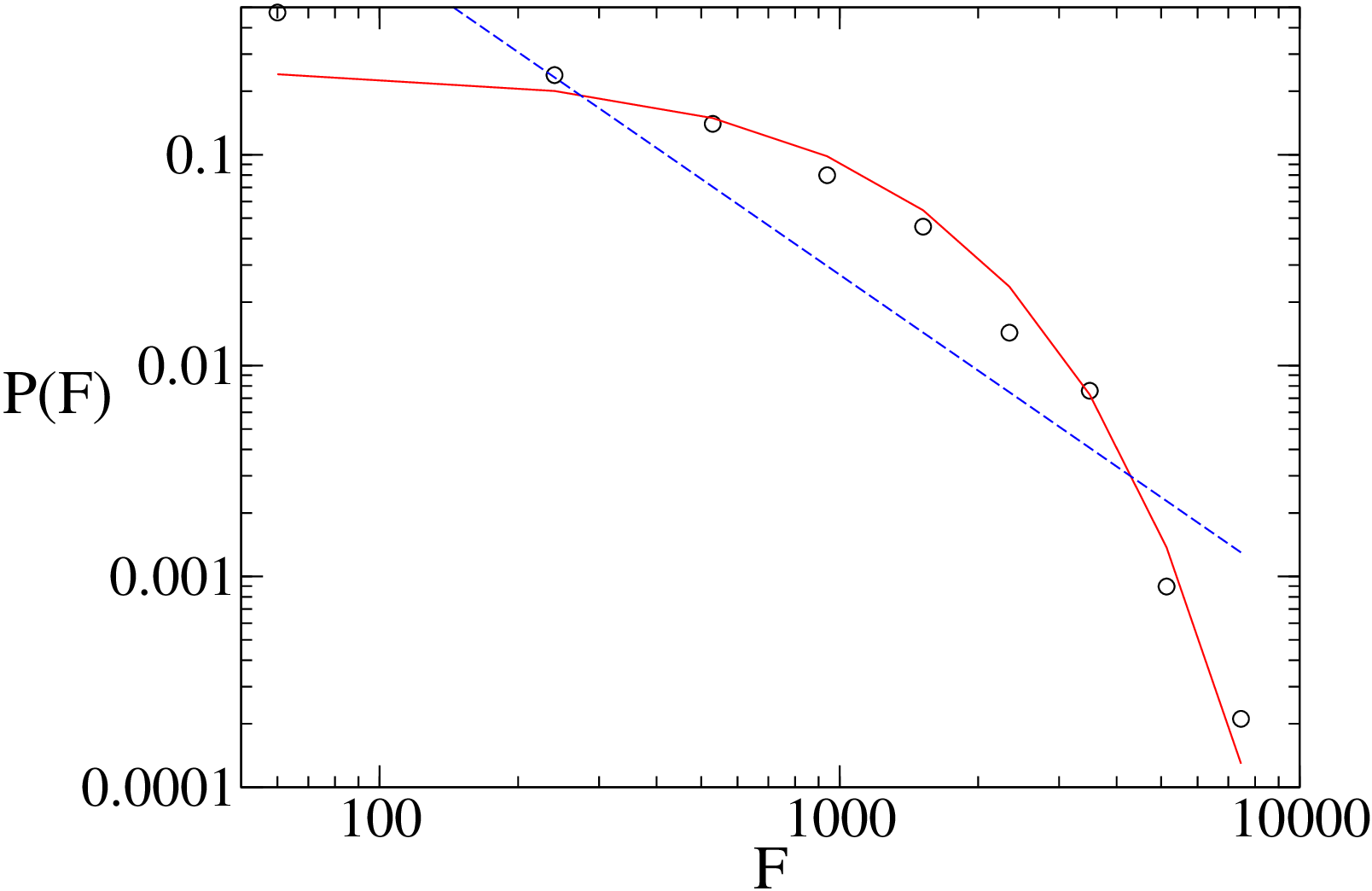}
\caption{Distribution of fame in science.
The distribution is clearly better fitted by an exponential, 	$P(F)\sim0.26e^{-0.001F}$ (curve, $R^2=0.98$), than by a power-law (straight line, $R^2=0.82$).  
\label{p(f)science} }
\end{figure}

Our findings for scientists --- researchers in the area of condensed matter and statistical physics --- are dramatically different.  
We have examined a list of 449 researchers, drawn randomly from among those who post articles on the web-based electronic board {\em http://www.arxiv.org/archive/cond-mat}.  As a measure of fame, we used the UCLA \Google-hit criterion, with search lexicon:  ``Author's name"  AND ``condensed matter" OR ``statistical physics" OR ``statistical mechanics".  The distribution of fame, as measured by the number of hits, decays exponentially, rather than in power-law fashion (Fig.~\ref{p(f)science}):
\begin{equation}
\label{p(f).eq.sci}
P(F)\sim e^{-\eta F};\qquad\eta=0.00102\pm0.00006.
\end{equation}
The figure results, as well as the computed correlation for the two possibilities ($R^2=0.977$ for the exponential, vs.\ $0.82$ for a power-law), leave little doubt as to which is the better fit.

\begin{figure}
\includegraphics[width=0.4\textwidth]{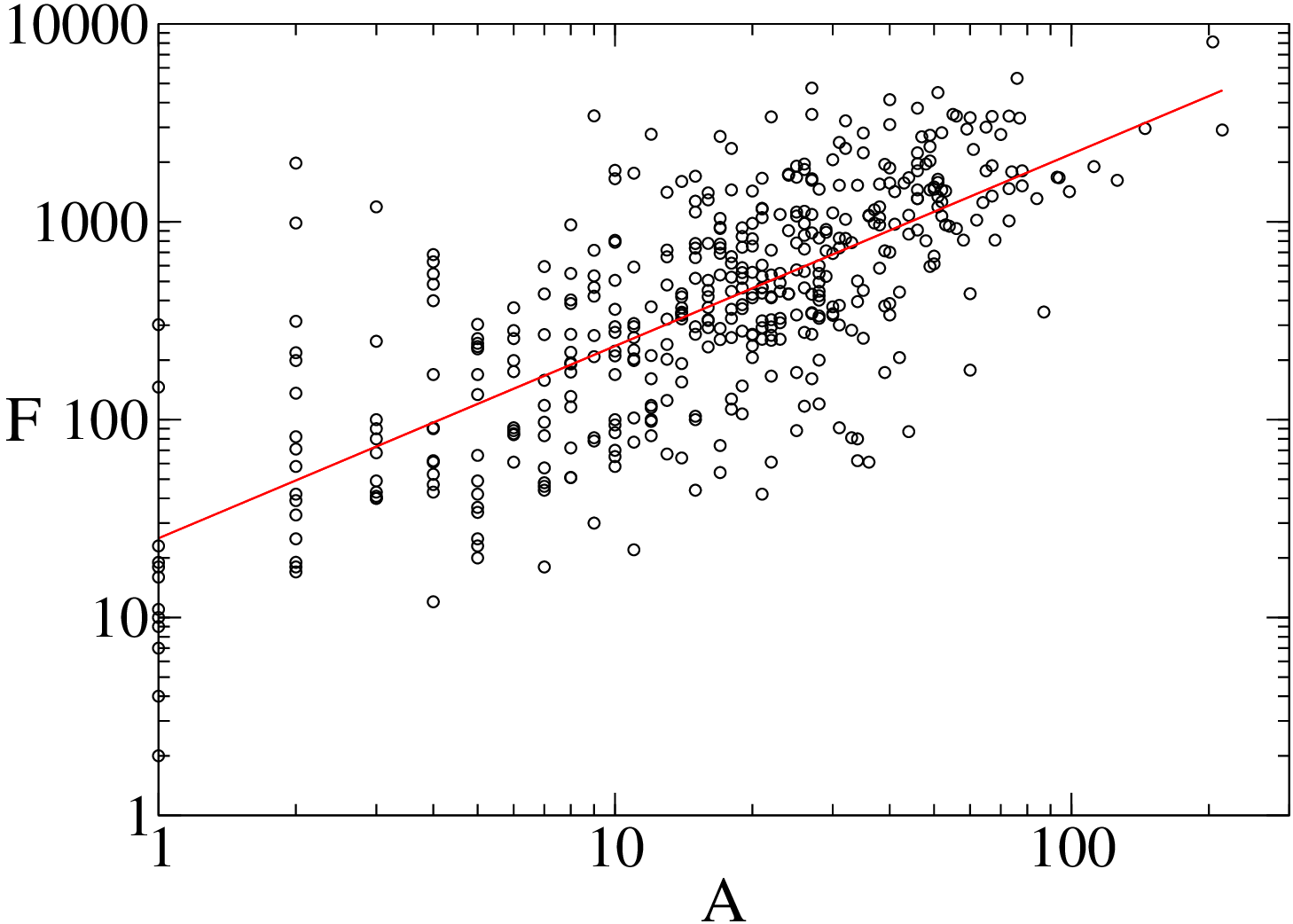}
\caption{ Fame of individual scientists (\Google\ hits) versus achievement (number of papers posted in the /cond-mat e-archive). The data ($\circ$) is better fitted by
an almost linear relation, $F\sim cA$ (straight line), than by the exponential dependence found in~\cite{Vwani}.
\label{f(a)science} }
\end{figure}

Next, we consider the relation between fame and achievement in science.  As a simple-minded measure of achievement we take the total number of publications by an author on the cond-mat board, perhaps thus paying too much heed to the popular  dictum ``publish or perish."  (The cond-mat board has been active since 1991.)  Our findings are summarized in Fig.~\ref{f(a)science}.  The wide scatter of data points resembles that found for ace pilots by the UCLA team, but our best fit indicates a power-law (almost linear) increase of fame with achievement, rather than the exponential dependence found there.  Indeed, we find $R^2=0.513$ for a power-law fit, vs.~$0.328$ for an exponential: quite the reverse from~\cite{Vwani}, which (for 393 ace pilots) cites $R^2=0.72$ for the exponential, vs.~$0.48$ for the linear fit.  Our findings suggest the linear relation:
\begin{equation}
\label{f(a).eq.sci}
F(A)\sim cA^{\xi};\qquad \xi=0.97\pm0.04\approx1.
\end{equation}

\begin{figure}
\includegraphics[width=0.4\textwidth]{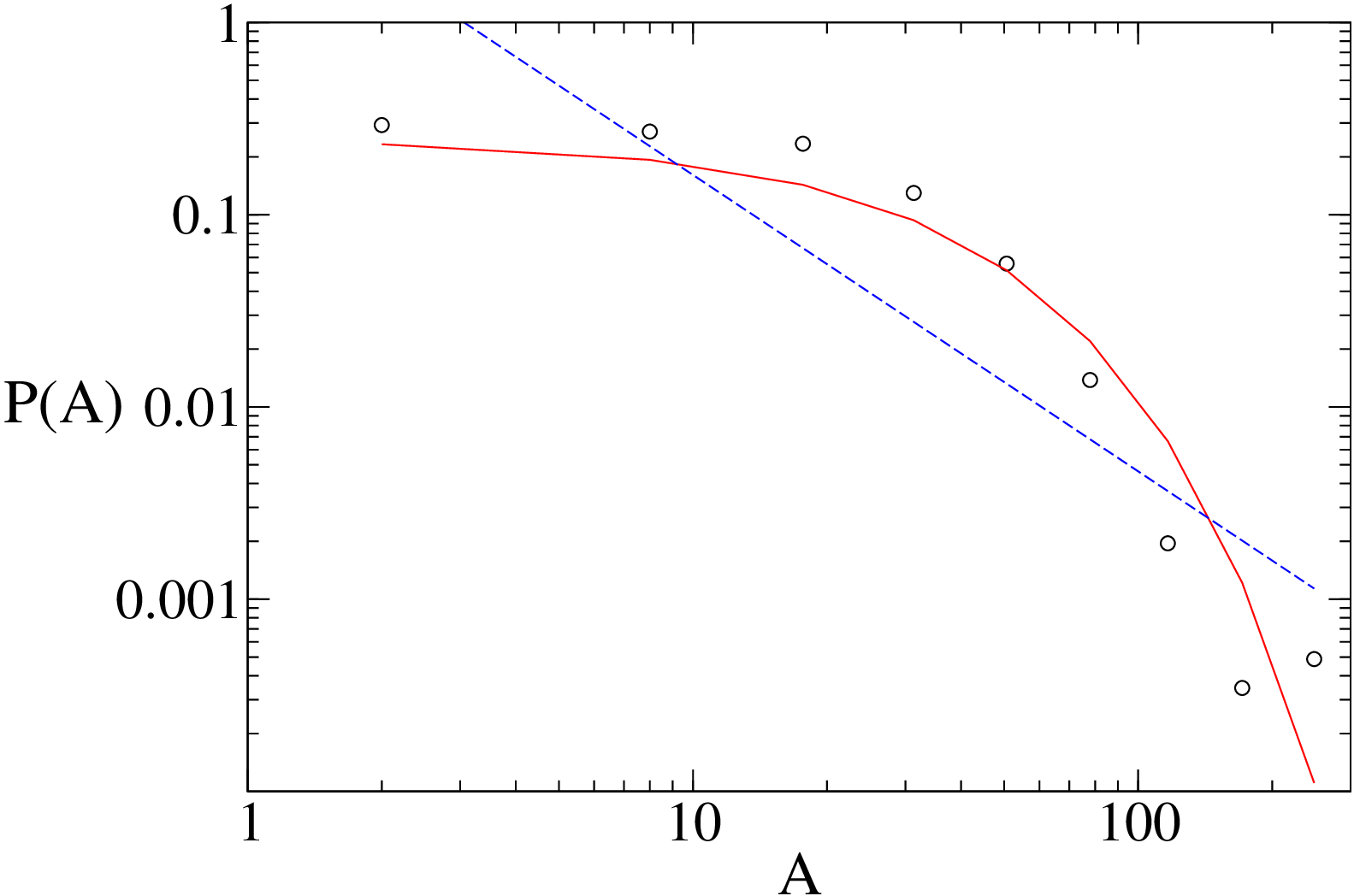}
\caption{ Distribution of achievement in science, as measured by the number of papers posted by each author on the web archive http://arxviv.org/archive/cond-mat.  The data ($\circ$) is better fitted by an exponential, $P(A)=0.25 e^{-0.031A}$ (curve, $R^2=0.90$), than by a power-law (straight line, $R^2=0.76$).
\label{p(a)science} }
\end{figure}

If Eqs.~(\ref{p(f).eq.sci}), (\ref{f(a).eq.sci}) are right (with $\xi=1$), it follows that achievement in science is distributed exponentially:
\begin{equation}
\label{p(a).eq.sci}
P(A)\sim e^{-\nu A};\qquad \nu=c\eta.
\end{equation}
Indeed, independent measurements of the probability distribution of achievement do support this prediction (Fig.~\ref{p(a)science}).
Moreover, the value found from a best fit for $\nu=0.031\pm0.004$ is consistent with $c=25\pm1$, $\eta=0.00102\pm0.00006$, and $c\eta=0.0255\pm0.0025$ found from the two previous plots.  We note that the observed exponential decay of the probability of achievement in science is the only feature that seems to be shared with the fame-achievement question in the case of ace pilots.

What could be the reason for the different fame vs.~merit patterns found for scientists and ace pilots?
One likely difference is in the set of people who author webpages that refer to ace pilots as opposed to those who write webpages about scientists.  In~\cite{Vwani} the authors explained their results using a
``rich-get-richer" scheme~\cite{Bara}, whereby individuals that are already popular attract people to generate new webpages at a rate proportional to their current popularity.  Implicit in this mechanism is the assumption that there is an inexhaustible (or at least very large) pool of people that may author webpages on a popular subject.  Such an assumption might perhaps be justified for a subject that enjoys wide notoriety within the public at large --- in other words, for truly famous subjects.  We maintain that, for whatever reason, scientists are simply not known to the general public, thereby curtailing the option of a rich-get-richer growth.  Instead, it is mostly scientist that write webpages about other scientists in their own discipline.  In fact, a simple explanation to the linear increase of fame (number of \Google\ hits) and achievement (number of papers published) in science is that scientists get cited on other scientists' webpages in relation to their published work.  If each published work typically generates citations in $c$ webpages, it follows that $A$ publications would connect a scientist with $F=cA$ webpages, on average.

\begin{figure}
\includegraphics[width=0.45\textwidth]{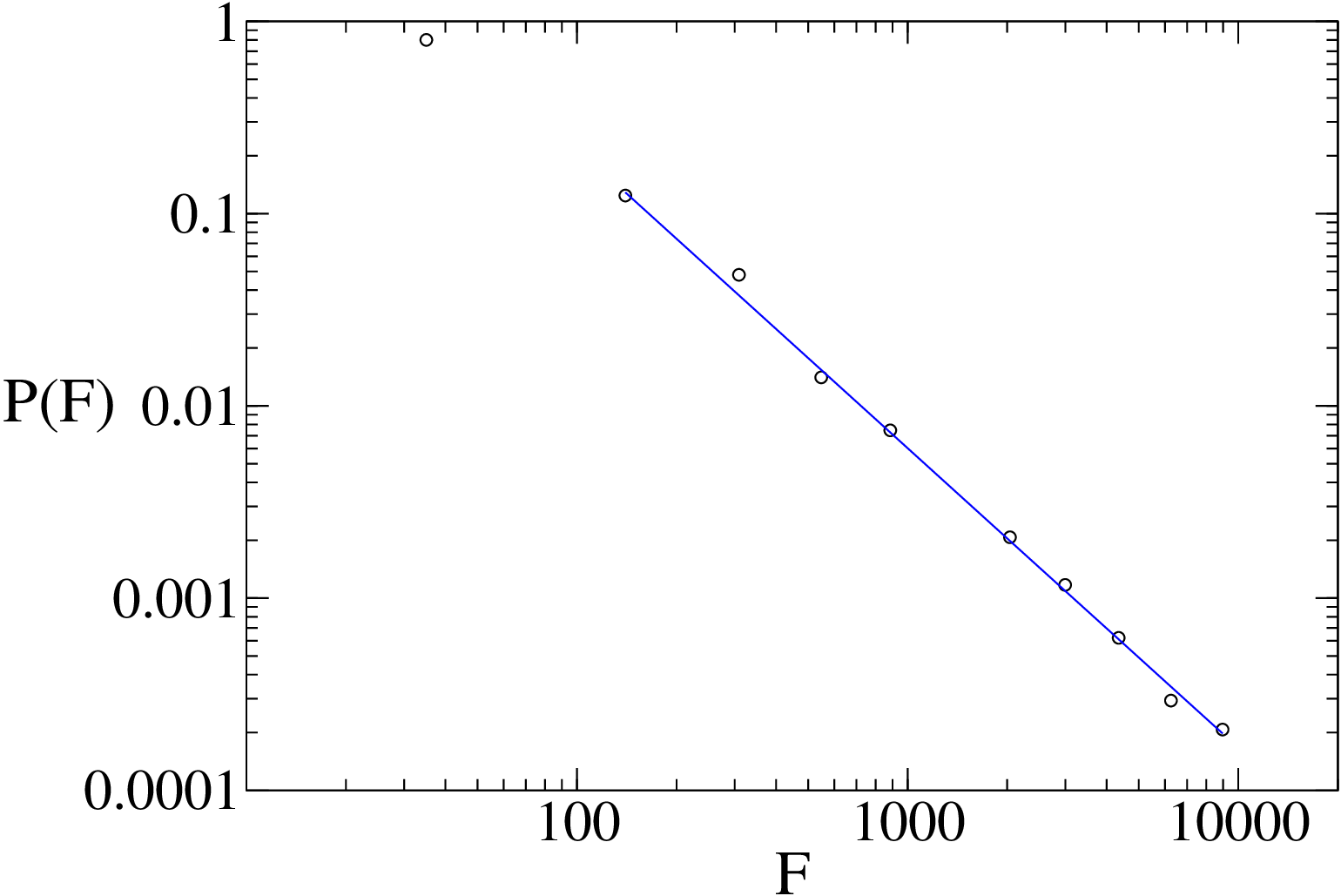}
\caption{Fame distribution for 291 athletes in various events:
100$\,$m, 200$\,$m, 1,500$\,$m, and 10,000$\,$m running, and 10,000$\,$m walking.
  The data ($\circ$) is well-fitted by a power-law (straight line) with exponent 1.55.
\label{runners} }
\vskip 0.15in
\end{figure}

\begin{figure}
\includegraphics[width=0.45\textwidth]{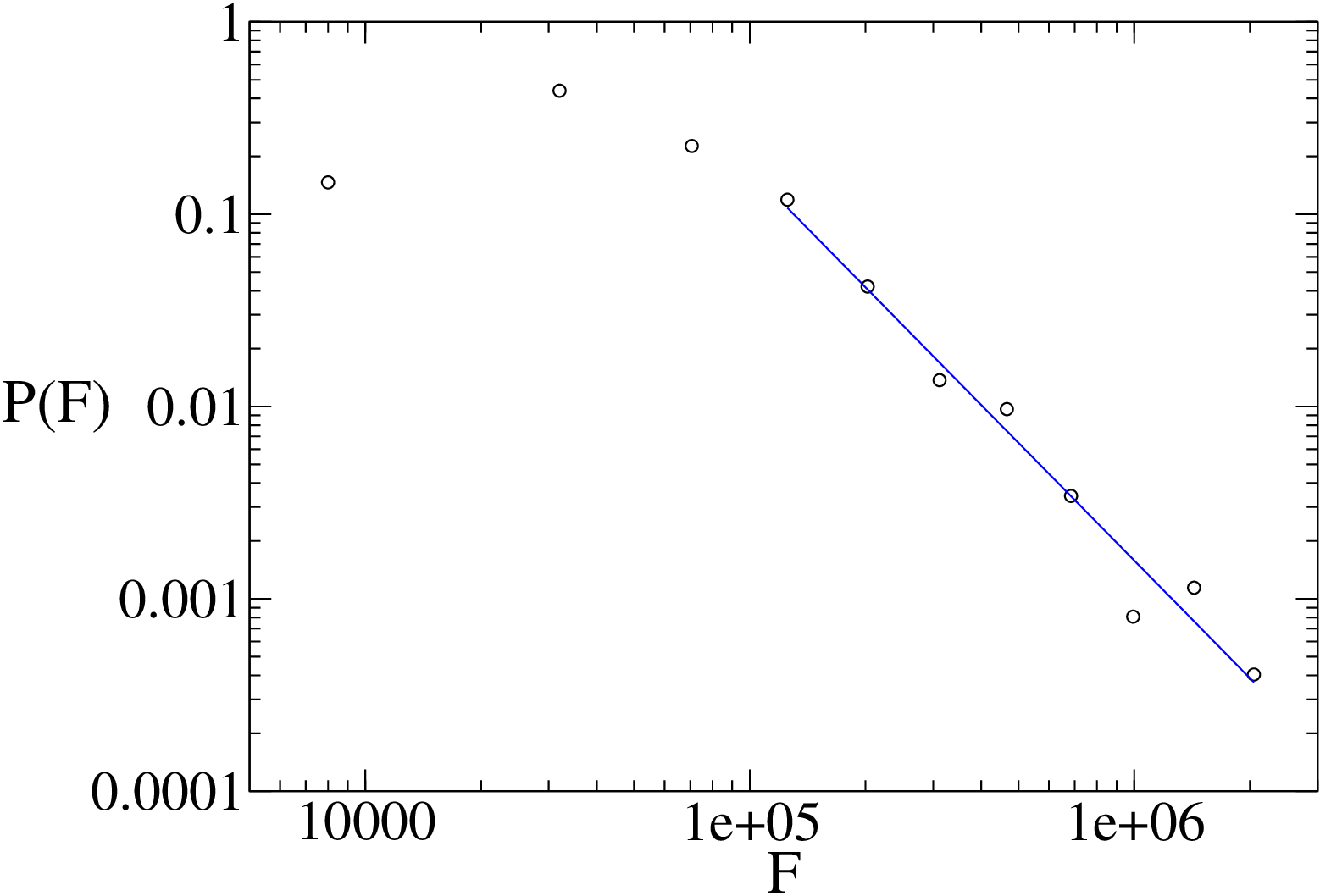}
\caption{ Fame distribution among 263 movies and television actors.  The data ($\circ$) is well fitted by a power-law (straight line) with exponent $2.03\pm0.14$, in line with the prediction in~\cite{Vwani}.
\label{actors} }
\vskip -0.15in
\end{figure}

It is also worth noting that it is not merely the relation between fame and merit that is different for scientists and ace pilots,  but also the distribution of fame itself.
This gives us a valuable simpler way to explore fame among various groups of people, for while achievement is not usually easy to define, one can always use the \Google-hit number as a universal measure of fame.   We expect that truly famous groups would exhibit a power-law distribution of fame, 
as for ace pilots, while those that are famous only within their own community might display a fast decaying or exponential distribution, as for scientists.  We have tested this idea against a group of 291 olympic runners (all distances) and a group of 263 television and movie actors (randomly drawn from an alphabetical list). The data, presented in Figs.~\ref{runners} and~\ref{actors} show that in both cases the probability of a certain level of fame drops off as a power-law, similar to Eq.~(\ref{fame_aces}), implying that  ``true" fame has been achieved --- fame within the public at large.  It is interesting, though, that the exponent $\gamma\approx1.55$ found for runners is quite different from $\gamma=2$ predicted by the theory advanced for ace pilots~\cite{Vwani}.  

The fame-merit behavior found for ace pilots is well documented at least in one other case, and independently from the \Google-hit measure of fame.
Consider a network of social contacts, where two individuals (nodes) are connected if they know each other on a first name basis.  In this case, the number of links $k$ coming out from a node constitutes an obvious measure of  fame, $F=k$.  As a measure of achievement, we may take the distance of a node from an individual at the top of the social ladder (the president of the USA, say).  A node is distance $l$ from the president if the shortest path connecting between the two, along the links of the net, consists of $l$ links.  Thus, people immediately connected to the president boast $l=1$, while those connected to them are (at least) distance $l=2$ away, etc.  We shall deem being {\em closer} to the president to represent a higher level of social achievement, that is $A=-l$.  It is well known that social nets of contacts are scale-free, possessing a degree distribution $P(k)\sim k^{-\gamma}$.  Hence,
\[
P(F)\sim F^{-\gamma},
\]
as in Eq.~(\ref{fame_aces}).  Meanwhile, since the number of nodes at distance $l$ increases as $\langle k \rangle^l$ (where $\langle k \rangle$ is the average degree of a node), we have
\[
P(A)\sim e^{-\alpha A}, \qquad \alpha=\ln\langle k\rangle.
\]
The two relations imply
\[
F(A)\sim e^{\beta} A,\qquad \beta=\alpha/(\gamma-1),
\]
exactly as found for ace pilots.  Note however that in this example the fame exponent $\gamma$
need not equal 2.  On the other hand, we know from recent studies on network growth that some
kind of rich-get-richer mechanism is invariably involved, whether implicitly or explicitly.

To summarize, the relations for fame and merit found for ace pilots are probably universal
among groups of people that enjoy true fame, within the general public.  Behind these relations there is a possible rich-get-richer mechanism for the acquisition of fame, though the details might vary from one group to the next.  Most importantly, the nature of fame can be studied on its own with the cheap and easy measure of number of \Google\ hits: power-law distributions indicate true fame.

For scientists, we find completely different relations, including an exponentially decaying distribution of fame.  This suggests that scientists (at least as measured by \Google\ hits) are primarily famous within their own science community, and do not enjoy public fame as other groups in the study do.
Needless to say, our study excluded the limited number of truly famous scientists, such as Albert Einstein (1,660,000 hits),
Isaac Newton (902,000), Galileo Galilei (245,000), Richard Feynman (112,000), and perhaps a handful of others.  Their popularity pales in comparison with other public figures, such as the performers Michael Jackson (5,570,000), Janet Jackson (3,190,000), and Barry Williams~\cite{greg} (2,400,000) .

Our findings for scientists might of course be affected by our choice of  the \Google-hit measure.
Had we measured fame in any other way, we may have found the telltale power-law distribution characteristic of other groups.  A reasonable argument is that scientists make heavy use of the WWW for propagation of their work, and as a result the number of  webpages dictated by their professional activities overwhelms the number of webpages written about them by the general public.
In this respect, there is comfort in our poor showing of fame, for at the same time it is an indication of a productive level of research~\cite{students}.

{}



\begin{thebibliography}{}
\bibitem{students} Indeed, a brief \Google\ search for undergraduate students at Clarkson University (lexicon: ``Student's Name" AND ``Clarkson University") --- another non-famous group --- shows that most students garner zero hits, and at most a few dozen hits, well below the least famous scientists in our study.
\end{thebibliography}
\end{document}